\documentclass[12pt,journal,onecolumn]{IEEEtran}
 
\IEEEoverridecommandlockouts

\usepackage{epsfig,rotating,setspace,latexsym,amsmath,epsf,amssymb,amsfonts,bm,theorem,cite,algorithm,graphicx,epsf,authblk,epstopdf,color,algpseudocode,bbm}

\newtheorem{remark}{Remark}

\allowdisplaybreaks

\begin{document}

\title{Secure Relaying in Non-Orthogonal Multiple Access: Trusted and Untrusted Scenarios
\thanks{Ahmed Arafa and H. Vincent Poor are with the Electrical Engineering Department at Princeton University, NJ 08544. Emails: \emph{aarafa@princeton.edu}, \emph{poor@princeton.edu}.}
\thanks{Wonjae Shin is with the Department of Electronics Engineering, Pusan National University, Busan, 46241, South Korea. He is also with the Electrical Engineering Department at Princeton University, NJ 08544. Email: \emph{wjshin@pusan.ac.kr}.}
\thanks{Mojtaba Vaezi is with the Electrical and Computer Engineering Department, Villanova University, PA 19085. Email: \emph{mvaezi@villanova.edu}.}
}

\author{Ahmed Arafa,~\IEEEmembership{Member,~IEEE}, Wonjae Shin,~\IEEEmembership{Member,~IEEE}, Mojtaba Vaezi,~\IEEEmembership{Senior Member,~IEEE}, and H. Vincent Poor,~\IEEEmembership{Fellow,~IEEE}}

\maketitle

%================================
\begin{abstract}
A downlink single-input single-output non-orthogonal multiple access setting is considered, in which a base station (BS) is communicating with two legitimate users in two possible scenarios of unsecure environments: existence of an external eavesdropper and communicating through an {\it untrusted} relay. For the first scenario, a number of {\it trusted} cooperative half-duplex relays is employed to assist with the BS's transmission and secure its signals from the external eavesdropper. Various relaying schemes are proposed and analyzed for that matter: {\it cooperative jamming}, {\it decode-and-forward}, and {\it amplify-and-forward}. For each scheme, {\it secure beamforming} signals are devised at the relays to maximize the achievable secrecy rate regions. For the second scenario, with the untrusted relay, achievable secrecy rate regions are derived for two different relaying schemes: {\it compress-and-forward} and {\it amplify-and-forward}, under two different modes of operation. In the first mode, coined {\it passive user} mode, the users receive signals from both the BS and the untrusted relay, and combine them to decode their messages. In the second mode, coined {\it active user} mode, the users transmit a cooperative jamming signal {\it simultaneously} with the BS's transmission to further confuse the relay. Focusing on half-duplex nodes, the users cannot receive the BS's signal while jamming the relay, i.e., while being active, and rely only on the signals forwarded to them by the relay. It is shown that the best relaying scheme highly depends on the system parameters, in particular distances between the nodes, and also on which part of the secrecy rate region the system is to operate at.
\end{abstract}

%================================
\section{Introduction}

Non-orthogonal multiple access (NOMA) techniques offer promising solutions to spectrum scarcity and congestion problems in next-generation wireless networks, attributed to its efficient utilization of available resources serving multiple users simultaneously, as opposed to conventional orthogonal multiple access techniques \cite{poor-noma-intro, mojtaba-multiple-access-5g}. Owing to the broadcast nature of wireless transmissions, securing transmitted data from potential eavesdroppers or untrusted nodes in the network is a critical system design aspect that needs careful consideration. Physical layer security is a powerful tool to achieve the goal of, a provably unbreakable, secure communications by exploiting the inherently different physical communication channels between different nodes in the network, see, e.g., \cite{poor-wireless-pls} and the references therein. In this work, we design secure transmission schemes for a downlink single-input single-output (SISO) NOMA system considering two possible unsecure environments: the first is when there is an external eavesdropper, for which we use {\it trusted cooperative} relays to enhance security, and the second is when communication occurs through an {\it untrusted} relay node.

There has been a number of recent works in the literature that study physical layer security for NOMA systems \cite{ding-noma-sum-sec-rate, hanzo-noma-sec-large-scale-downlink, elkashlan-noma-sec-large-scale-uplink, ding-noma-unicast-multicast, qin-noma-sum-sec-eav-mimo, qin-noma-sum-sec-edge-miso, qin-noma-sum-sec-edge-mimo, alouini-noma-sec-antenna-select, lau-noma-outage}. Secrecy sum rate maximization of SISO NOMA systems is studied in \cite{ding-noma-sum-sec-rate}. Using tools from stochastic geometry, references \cite{hanzo-noma-sec-large-scale-downlink} and \cite{elkashlan-noma-sec-large-scale-uplink} study security measures for large-scale NOMA systems in the downlink and the uplink, respectively. NOMA assisted multicast-unicast streaming is studied in \cite{ding-noma-unicast-multicast}, where secure rates for unicast transmission using NOMA is shown to outperform conventional orthogonal schemes. Reference \cite{qin-noma-sum-sec-eav-mimo} considers a multiple-input multiple-output (MIMO) two-user NOMA setting with an external eavesdropper and designs beamforming signals that maximize the secrecy sum rate. This approach is also considered in \cite{qin-noma-sum-sec-edge-miso} and \cite{qin-noma-sum-sec-edge-mimo} for multiple-input single output (MISO) and MIMO scenarios, respectively, in a two-user setting, with the assumption that one user is entrusted and the other is the potential eavesdropper. The impacts of transmit antenna selection strategies on the secrecy outage probability is investigated in \cite{alouini-noma-sec-antenna-select}. Transmit power minimization and minimum secrecy rate maximization subject to a secrecy outage constraint are considered in \cite{lau-noma-outage}. Different from the previous works, in this work we investigate the advantages of using trusted cooperative relays to secure messages from an external eavesdropper, and also study the impact of having an untrusted relay on achievable secrecy rates in the context of NOMA.

Our work on using trusted cooperative relays is most closely related to the single-receiver wiretap channel work in \cite{petropulu-coop-relay-security}, in which half-duplex relays are employed to enhance security. Reference \cite{raef-df-cj} uses similar ideas with the focus on full-duplex relays using mixed decode-and-forward and cooperative jamming strategies. Information-theoretic analysis of communication systems with untrusted relay nodes are considered in \cite{lai-elgamal-relay-eav, raef-deaf-relay-select, raef-deaf-relay-mult-antenna, he-yener-two-hop-untrusted, zewail-untrusted-2hop-multi-receiver, he-yener-untrusted-relay}. References \cite{lai-elgamal-relay-eav, raef-deaf-relay-select, raef-deaf-relay-mult-antenna} consider the setting of deaf relays, i.e., relays that are ignorant of the source's transmitted signal, in the presence of external eavesdroppers, and develop achievable secrecy rates based on cooperative jamming and noise forwarding schemes. References \cite{he-yener-two-hop-untrusted} and \cite{zewail-untrusted-2hop-multi-receiver} study a two-hop scenario with an untrusted relay (with no external eavesdroppers) and provide achievable secrecy rates for a single source-destination pair and for a multi terminal setting, respectively, with the help of cooperative jamming signals from the destination(s). The general untrusted relay channel (also with no external eavesdropper) is considered in \cite{he-yener-untrusted-relay}, where positive secrecy rates are shown achievable if the source-relay channel is orthogonal to the relay-destination channel via compress-and-forward scheme at the relay. For a summary of cooperative security works, see, e.g., \cite{ulukus-coop-sec-summary} and the references therein. Similar to the previous references, in this work we also use information-theoretic tools to derive achievable secrecy rate regions in the context of NOMA, with trusted and untrusted relays.

In the first part of this paper, we extend the ideas in \cite{petropulu-coop-relay-security} to work in the context of a two-user downlink SISO NOMA system with an external eavesdropper. We employ multiple {\it trusted cooperative} half-duplex relays to enhance the achievable secrecy rate region through various relaying schemes: {\it cooperative jamming}, {\it decode-and-forward}, and {\it amplify-and-forward}. For each scheme, we design {\it secure beamforming} signals at the relays that benefit the users and/or hurt the eavesdropper. Under a total system power constraint, that is divided between the base station (BS) and the relays, an achievable secrecy rate region for each relaying scheme is derived and analyzed. In general, the results in this case show that the best relaying scheme highly depends on the system parameters, in particular the distances between nodes, and that the relatively simple cooperative jamming scheme performs better than the other schemes when the relays are close to the eavesdropper.

In the second part of this paper, we consider a different scenario in which an {\it untrusted} half-duplex relay node is available to assist with the BS's transmission. Applications of this scenario are when, e.g., the relay has a lower security clearance relative to the end users, and hence transmission schemes should be designed in such a way that the relay can only forward the data without revealing its actual contents. We derive achievable secrecy rate regions in this case under two relaying schemes: {\it compress-and-forward} and {\it amplify-and-forward}. We also consider two modes of operations: {\it passive user} mode and {\it active user} mode. In the passive user mode, the users receive data from both the BS and the relay and combine them efficiently to decode their messages, while in the active user mode, the users transmit a cooperative jamming signal simultaneously with the BS's transmission to further confuse the relay, and hence, since the focus is on half-duplex nodes, they cannot receive the BS's transmission and rely solely on the data forwarded to them through the relay. We derive, analyze, and compare the achievable secrecy rate regions for each relaying scheme and operating mode under a total system power constraint, that is divided between the BS, the relay, and the users if operating in the active mode. As in the first part of the paper, the results also show in this case that the best relaying scheme and operating mode depends, in particular, on the distances between the nodes, with a general superiority of the active user mode over the passive user mode.

%================================
\section{System Model}

We consider a downlink NOMA system where a BS is communicating with two users. All nodes, including the BS, are equipped with single antennas. Channels from the BS to the users are fixed during the communication session, and are known at the BS. In a typical NOMA downlink setting, the BS uses superposition coding to send messages to the two users simultaneously. The user with a relatively worse channel condition (weak user) decodes its message by treating the other user's interfering signal as noise, while the user with a relatively better channel condition (strong user) first decodes the weak user's message, by treating its own interfering signal as noise, and then uses successive interference cancellation to decode its own message.\footnote{The two-user setting in this work is adopted in NOMA systems in which users are divided into multiple clusters with two users each, in order to reduce error propagation in successive interference cancellation decoding \cite{poor-noma-intro}.}

All channel gains in this paper are complex-valued, and are drawn independently from some continuous distribution. We denote the channel between the BS and the strong user (resp. weak user) by $h_1$ (resp. $h_2$), with $|h_1|^2\geq|h_2|^2$. The channel gains $h_1$ and $h_2$ are known at the BS. The received signals at the strong and weak users are given by
\begin{align}
y_1&=h_1x+n_1, \label{eq_rec_sig_1} \\
y_2&=h_2x+n_2, \label{eq_rec_sig_2}
\end{align}
where the noise terms $n_1$ and $n_2$ are independent and identically distributed (i.i.d.) circularly-symmetric complex Gaussian random variables with zero mean and unit variance, $\mathcal{CN}\left(0,1\right)$, and the transmitted signal $x$ is given by
\begin{align} \label{eq_trans_sgnl}
x=\sqrt{\alpha P}s_1+\sqrt{\bar{\alpha}P}s_2,
\end{align}
where $s_1$ and $s_2$ are i.i.d. $\sim\mathcal{CN}\left(0,1\right)$ information carrying signals for the strong and the weak user, respectively, $P$ is the BS's transmit power budget, $\alpha\in[0,1]$ is the fraction of power allocated to the strong user, and $\bar{\alpha}\triangleq1-\alpha$. Using superposition coding and successive interference cancellation decoding, one achieves the following rates of this (degraded) Gaussian broadcast channel \cite{cover}:\footnote{The $\log$ terms in this paper denote natural logarithms.}
\begin{align}
r_1&=\log\left(1+|h_1|^2\alpha P\right), \label{eq_rate_1}\\
r_2&=\log\left(1+\frac{|h_2|^2\bar{\alpha}P}{1+|h_2|^2\alpha P}\right).\label{eq_rate_2}
\end{align}

In the next sections, we discuss several relaying schemes to deliver secure data to both users. Specifically, in Section~\ref{sec_trstd}, we investigate the case in which multiple cooperative trusted relays help securing the data from an external eavesdropper, and then we investigate the case of having an untrusted relay in Section~\ref{sec_untrstd}.

%================================
\section{Trusted Relays with an External Eavesdropper} \label{sec_trstd}

In this section, we consider the situation in which there is an external eavesdropper that is monitoring the communication between the BS and the users. We denote the channel between the BS and the eavesdropper by $h_e$, and assume that it is known at the BS. The received signal at the eavesdropper is given by
\begin{align}
y_e=h_ex+n_e, \label{eq_rec_sig_e}
\end{align}
where the noise term $n_e\sim\mathcal{CN}\left(0,1\right)$. For a given $0\leq\alpha\leq1$, the secrecy capacities of the two users in this multi-receiver wiretap channel are given by \cite[Theorem 5]{ersen-mr-wt}
\begin{align}
r_{s,1}&\!=\!\left[\log\left(1+|h_1|^2\alpha P\right)-\log\left(1+|h_e|^2\alpha P\right)\right]^+\!\!, \label{eq_sec_rate_1} \\
r_{s,2}&\!=\!\left[\log\left(\!1+\frac{|h_2|^2\bar{\alpha}P}{1+|h_2|^2\alpha P}\!\right)-\log\left(\!1+\frac{|h_e|^2\bar{\alpha}P}{1+|h_e|^2\alpha P}\!\right)\right]^+\!\!, \label{eq_sec_rate_2}
\end{align}
where the subscript $s$, here and throughout the paper, is to denote secrecy rates, and $[x]^+\triangleq\max\{x,0\}$.

\begin{figure}[t]
\center
\includegraphics[scale=1]{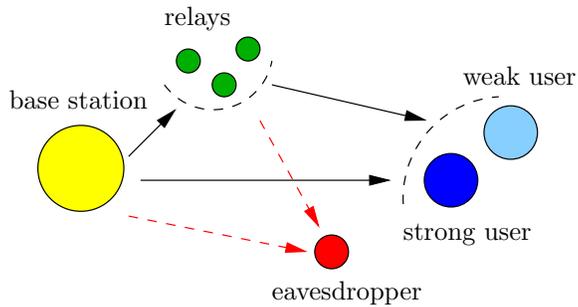}
\caption{Downlink NOMA system model, with cooperative relays, and an external eavesdropper.}
\label{fig_sys_model_trstd}
\end{figure}

It is clear from (\ref{eq_sec_rate_1}) and (\ref{eq_sec_rate_2}) that sending secure data depends on the eavesdropper's channel condition with respect to that of the legitimate users. Therefore, we propose using trusted cooperative half-duplex {\it relay} nodes, see Fig.~\ref{fig_sys_model_trstd}, to assist the BS via three possible schemes: {\it cooperative jamming}, {\it decode-and-forward}, and {\it amplify-and-forward}. In all of these schemes, the BS uses only a portion of its available power $\bar{P}\leq P$ for its own transmission, and shares the remaining portion $P-\bar{P}$ with the relays for their transmission. The main reason behind such power reduction at the BS is to have a fair comparison between relaying and non-relaying scenarios. This way, the value of $P$ represents a system's {\it total} power budget that is to be distributed among its different transmitting nodes. We note that such approach has been adopted in the single-user setting in \cite{petropulu-coop-relay-security}, and that without it, one would have the (unrealistic) ability to add beneficial relay nodes at no additional cost. Our numerical results in Section~\ref{sec_num}, however, show that under a total system's power budget, some relaying schemes might not be that helpful in situations where relays are relatively far away from the BS. We discuss the relaying schemes in details over the next subsections. In what follows, we introduce the relays' channels notation that we use.

Let there be $K$ relays, and denote the channel gains from the BS to the relays by the vector ${\bm h}_r\triangleq[h_{r,1},\dots,h_{r,K}]$.\footnote{All vectors in this paper are column vectors. For instance, ${\bm h}_r$ is a $K\times1$ vector.} Let ${\bm g}_1$, ${\bm g}_2$, and ${\bm g}_e$ denote the $K$-length channel gain vectors from the relays to the first user, the second user, and the eavesdropper, respectively. We assume that $|g_{1,k}|^2\geq|g_{2,k}|^2$, $1\leq k\leq K$. That is, the strong user with respect to the BS is also strong with respect to the $k$th relay. This is satisfied, for instance, in the typical scenario in which the relays are closer to the BS than both legitimate users. The relays are {\it cooperative} in the sense that they design their transmission schemes based upon sharing knowledge of their channel state information (CSI) among themselves, as in \cite{petropulu-coop-relay-security}. The channels from the relays to the users and the eavesdropper are known at the relays, and at the BS. 

We note that assuming perfect knowledge of the external eavesdropper's CSI at the legitimate transmitters enables developing a fundamental understanding of how cooperation can enhance security in NOMA settings by characterizing achievable secrecy rate regions. Other situations in which only partial/statistical knowledge of external eavesdroppers' CSI is available have been considered in the physical layer security literature through providing worst case security guarantees via, e.g., characterizing secrecy outage probabilities, see \cite{poor-wireless-pls} and the references therein. This is different from the approach considered in this paper, and is delegated to future works on the subject.

\subsection{Cooperative Jamming} \label{sec_coop_jam}

In this subsection, we discuss the cooperative jamming scheme. {\it Simultaneously} with the BS's transmission, the relays transmit an artificial noise/cooperative jamming signal ${\bm J}z$, where ${\bm J}\in\mathbb{C}^K$ is a beamforming vector and $z\sim\mathcal{CN}(0,1)$, and hence the received signals at the users and the eavesdropper are in this case are given by
\begin{align}
y_1&=h_1x+{\bm g}_1^\dagger{\bm J}z+n_1, \\
y_2&=h_2x+{\bm g}_2^\dagger{\bm J}z+n_2, \\
y_e&=h_ex+{\bm g}_e^\dagger{\bm J}z+n_e,
\end{align}
where the superscript $\dagger$ denotes the conjugate transpose operation. The signal power is now set to $\mathbb{E}\left[|x|^2\right]=\bar{P}\leq P$, where $\mathbb{E}[\cdot]$ is the expectation operator, and the remaining portion $P-\bar{P}$ is used to power the relays, i.e., $\mathbb{E}\left[\left({\bm J}z\right)^\dagger\left({\bm J}z\right)\right]={\bm J}^\dagger{\bm J}=P-\bar{P}$.

Designing the jamming signal is such that it has minimal effect on the legitimate users as follows:
\begin{align} \label{eq_cj_null}
\left[{\bm g}_1 ~~~ {\bm g}_2\right]^\dagger{\bm J}\triangleq{\bm G}^\dagger{\bm J}={\bm 0}_{(2)},%\left[0~~~0\right]{\color{red}^\dagger}.%{\color{red}\begin{bmatrix}0\\0\end{bmatrix}}.
\end{align}
where ${\bm 0}_{(2)}$ denotes an all-zero (column) vector of size $2$. Thus, we choose ${\bm J}$ in the null space of the matrix ${\bm G}^\dagger$. If there exist $K\geq3$ relays, then ${\bm G}^\dagger$ will always have a nonempty null space and (\ref{eq_cj_null}) will have a nontrivial solution. We denote such null space jamming signal by ${\bm J}_o$. Since the channel state vectors ${\bm g}_1$, ${\bm g}_2$, and ${\bm g}_e$ are drawn from a continuous distribution, they are therefore linearly independent with probability $1$ (w.p. $1$). Thus, we have
\begin{align}
\left|{\bm g}_e^\dagger{\bm J}_o\right|>0,\quad\text{w.p. $1$}.
\end{align}
Therefore, the achievable secrecy rates are now given by (\ref{eq_sec_rate_1}) and (\ref{eq_sec_rate_2}) after replacing $h_e$ by $\tilde{h}_e$ defined as
\begin{align} \label{eq_he_tilde}
\tilde{h}_e\triangleq h_e\Big/\left(1+\left|{\bm g}_e^\dagger{\bm J}_o\right|^2\right).
\end{align}

We now find the optimal ${\bm J}_o$ that maximally degrades the eavesdropper's channel (subject to not affecting the legitimate users' channels). Upon replacing $h_e$ in (\ref{eq_sec_rate_1}) and (\ref{eq_sec_rate_2}) by $\tilde{h}_e$ defined above, one can directly see that both legitimate users' secrecy rates are increasing in $\left|{\bm g}_e^\dagger{\bm J}_o\right|^2$. Therefore, to maximize them, we formulate the following optimization problem for a given transmit power $\bar{P}$:
\begin{align}
\max_{{\bm J}_o} \quad &\left|{\bm g}_e^\dagger{\bm J}_o\right|^2 \nonumber \\
\mbox{s.t.} \quad &{\bm G}^\dagger{\bm J}_o={\bm 0}_{(2)} \nonumber \\%\left[0~~~0\right]{\color{red}^\dagger}%{\color{red}\begin{bmatrix}0\\0\end{bmatrix}}. \nonumber \\
&{\bm J}_o^\dagger{\bm J}_o=P-\bar{P}.
\end{align}
The above problem has a unique solution \cite{friedlander-null-steering} (see also \cite{petropulu-coop-relay-security}), which we derive next for completeness. We first rewrite the first constraint slightly differently as follows:
\begin{align}
{\bm J}_o=\mathcal{P}^\perp\!\left({\bm G}\right){\bm u}_J,
\end{align}
for some vector ${\bm u}_J\in\mathbb{C}^K$ to be designed, and $\mathcal{P}^\perp\!\left({\bm G}\right)$ is the orthogonal projection matrix onto the null space of ${\bm G}^\dagger$ given by
\begin{align} \label{eq_p_perp_original}
\mathcal{P}^\perp\!\left({\bm G}\right)\triangleq{\bm I}_K-{\bm G}\left({\bm G}^\dagger{\bm G}\right)^{-1}{\bm G}^\dagger,
\end{align}
where ${\bm I}_K$ is the $K$-dimensional identity matrix. It is now direct to see that the vector ${\bm u}_J$ should be chosen along the same direction of $\mathcal{P}^\perp\!\left({\bm G}\right){\bm g}_e$ in order to maximize $\left|{\bm g}_e^\dagger{\bm J}_o\right|^2$. Finally, to satisfy the power constraint, the optimal beamforming vector, $\hat{{\bm J}}_o$, is given by
\begin{align}
\hat{{\bm J}}_o=\frac{\mathcal{P}^\perp\!\left({\bm G}\right){\bm g}_e}{\left\|\mathcal{P}^\perp\!\left({\bm G}\right){\bm g}_e\right\|}\sqrt{P-\bar{P}},
\end{align}
which, upon substituting in (\ref{eq_he_tilde}), achieves the following secrecy rates for a given transmit power $\bar{P}$ and power fraction $\alpha$:
\begin{align}
r_{s,1}^J\!&=\!\left[\log\!\left(1+|h_1|^2\alpha \bar{P}\right)\!-\!\log\!\left(1+\frac{|h_e|^2\alpha \bar{P}}{1+{\bm g}_e^\dagger\mathcal{P}^\perp\!\left({\bm G}\right){\bm g}_e\left(P-\bar{P}\right)}\right)\!\right]^+, \\
r_{s,2}^J\!&=\!\left[\log\left(1+\frac{|h_2|^2\bar{\alpha}\bar{P}}{1+|h_2|^2\alpha \bar{P}}\right)-\log\left(1+\frac{|h_e|^2\bar{\alpha}\bar{P}}{1\!+\!|h_e|^2\alpha \bar{P}\!+\!{\bm g}_e^\dagger\mathcal{P}^\perp\!\left({\bm G}\right){\bm g}_e\left(P-\bar{P}\right)}\right)\right]^+, 
\end{align}
where the superscript $J$ is to denote the cooperative jamming scheme.

In Section~\ref{sec_num}, we discuss the evaluation of the optimal transmit power $\bar{P}$ and the power fraction $\alpha$ that maximize the secrecy rate region of this cooperative jamming scheme, along with those that maximize the secrecy rate regions of the other relaying schemes that we consider in the upcoming subsections.

\subsection{Decode-and-Forward} \label{sec_dec_fwd}

In this subsection, we discuss the decode-and-forward scheme. Different from cooperative jamming, communication takes place in the decode-and-forward scheme over two phases. In the first phase, the BS broadcasts the messages to both the relays and the legitimate users. In the second phase, the relays forward the messages that they decoded to the legitimate users. The eavesdropper overhears the communication during both phases. 

The received signals during the first phase at the legitimate users and the eavesdropper are given by (\ref{eq_rec_sig_1})--(\ref{eq_rec_sig_2}) and (\ref{eq_rec_sig_e}), respectively, with a total transmit power $\bar{P}\leq P$. The received signals at the relays during the first phase are given by
\begin{align} \label{eq_rec_sig_r}
{\bm y}_r={\bm h}_rx+{\bm n}_r,
\end{align}
where the noise term vector ${\bm n}_r\sim\mathcal{CN}\left(0,{\bm I}_K\right)$. During the first phase, each relay first decodes the weak user's message by treating the strong user's interfering signal as noise, and then uses successive interference cancellation to decode the strong user's message. Thus, the achievable rates at the $k$th relay after the first phase are
\begin{align}
R_{k,1}&=\log\left(1+|h_{r,k}|^2\alpha\bar{P}\right) \\
R_{k,2}&=\log\left(1+\frac{|h_{r,k}|^2\bar{\alpha}\bar{P}}{1+|h_{r,k}|^2\alpha\bar{P}}\right)
\end{align}

In the second phase, the relays form the transmitted signal $x_r$, which is exactly as in (\ref{eq_trans_sgnl}) but after replacing $P$ with $P-\bar{P}$. We assume that the relays use the same power fraction $\alpha$ in the second phase. While in general each relay can use a different power fraction, we use the same fraction for simplicity of presentation of the scheme hereafter. The relays use a unit-norm beamforming vector ${\bm d}\in\mathbb{C}^K$ during the second phase, to be designed, i.e., the $k$th relay multiplies the transmitted signal $x_r$ by $d_k$ and sends it through the channel, and hence the received signals at the legitimate users and the eavesdropper are given by
\begin{align}
y^r_1&={\bm g_1}^\dagger{\bm d}x_r+n^r_1,\\
y^r_2&={\bm g_2}^\dagger{\bm d}x_r+n^r_2,\\
y^r_e&={\bm g_e}^\dagger{\bm d}x_r+n^r_e,
\end{align}
where the superscript $r$ is to denote signals received from the relays, and the noise terms $n^r_1$, $n^r_2$, and $n^r_e$ are i.i.d. $\sim\mathcal{CN}\left(0,1\right)$. We let the relays use independent codewords from those used by the BS to forward their messages. Therefore, the achieved rates at the legitimate users after the second phase are given by (\ref{eq_dec_rate_1}) and (\ref{eq_dec_rate_2}) at the top of next page \cite[Theorem 16.2]{elgamalKim}, with the superscript $DF$ denoting decode-and-forward.
\begin{figure*}[t]
\begin{align}
%r^{DF}_1&=\min\left\{\log\left(1+|h_1|^2\alpha\bar{P}\right)+\log\left(1+\left|{\bm g}_1^\dagger{\bm d}\right|^2\alpha\left(P-\bar{P}\right)\right),\min_{1\leq k\leq K}R_{k,1}^{(i)}\right\} \label{eq_dec_rate_1} \\
%r^{DF}_2&=\min\left\{\log\left(1+\frac{|h_2|^2\bar{\alpha}\bar{P}}{1+|h_2|^2\alpha \bar{P}}\right)+\log\left(1+\frac{\left|{\bm g}_2^\dagger{\bm d}\right|^2\bar{\alpha}\left(P-\bar{P}\right)}{1+\left|{\bm g}_2^\dagger{\bm d}\right|^2\alpha \left(P-\bar{P}\right)}\right),\min_{1\leq k\leq K}R_{k,2}^{(i)}\right\} \label{eq_dec_rate_2}
r^{DF}_1&=\min\left\{\log\left(1+|h_1|^2\alpha\bar{P}\right)+\log\left(1+\left|{\bm g}_1^\dagger{\bm d}\right|^2\alpha\left(P-\bar{P}\right)\right),\min_{1\leq k\leq K}R_{k,1}\right\} \label{eq_dec_rate_1} \\
r^{DF}_2&=\min\left\{\log\left(1+\frac{|h_2|^2\bar{\alpha}\bar{P}}{1+|h_2|^2\alpha \bar{P}}\right)+\log\left(1+\frac{\left|{\bm g}_2^\dagger{\bm d}\right|^2\bar{\alpha}\left(P-\bar{P}\right)}{1+\left|{\bm g}_2^\dagger{\bm d}\right|^2\alpha \left(P-\bar{P}\right)}\right),\min_{1\leq k\leq K}R_{k,2}\right\} \label{eq_dec_rate_2}
\end{align}
\hrulefill
\end{figure*}

The first terms in the two minima in (\ref{eq_dec_rate_1}) and (\ref{eq_dec_rate_2}) represent the rates achieved through combining the signals from the BS and the relays at the users, while the second terms bind them by the achievable rates at the relays (from the BS), i.e., by the relays' ability to decode. Observe that since $|g_{1,k}|^2\geq|g_{2,k}|^2$, $1\leq k\leq K$, it follows that $\left|{\bm g}_1^\dagger{\bm d}\right|^2\geq\left|{\bm g}_2^\dagger{\bm d}\right|^2$, $\forall {\bm d}\in\mathbb{C}^K$. This ensures the ability of the strong user to decode the weak user's forwarded message from the relays successfully before employing successive interference cancellation.

We note that if the relays were to use the same codewords as those used by the BS, then one can view the whole system as a single-input multiple-output (SIMO) system at each user, wherein the first term in the minimum in (\ref{eq_dec_rate_1}) would slightly change to $\log\left(1+|h_1|^2\alpha\bar{P}+\left|{\bm g}_1^\dagger{\bm d}\right|^2\alpha\left(P-\bar{P}\right)\right)$, i.e., the signal-to-noise ratios (SNRs) at the strong user get added up, representing the SIMO capacity, see, e.g., the approach in \cite{petropulu-coop-relay-security}. Similar changes would also occur to the first term in the minimum in (\ref{eq_dec_rate_2}). We note that whether independent codewords or the same codewords are to be used at the relays, the approach we follow in the sequel to design the optimal beamforming vector ${\bm d}$ would not change. We choose, however, to continue with the independent codewords assumption as it achieves rates that are no smaller than those achieved via using the same codewords.

For $K\geq2$, we design the beamforming vector ${\bm d}$ to be a unit-norm vector orthogonal to ${\bm g}_e$, and denote it by ${\bm d}_o$. This way, the eavesdropper does not gain any useful information during the second phase. Thus, we have
\begin{align} \label{eq_dec_null}
{\bm g}_e^\dagger{\bm d}_o=0.
\end{align}
Further, for $K\geq3$, we have that $\{{\bm g}_1,~{\bm g}_2,~{\bm g}_e\}$ are linearly independent w.p. $1$, and hence
\begin{align}
|{\bm g}_1^\dagger{\bm d}_o|>0,~|{\bm g}_2^\dagger{\bm d}_o|>0,\quad\text{w.p. $1$}.
\end{align}
Thus, the achievable secrecy rates in this case are given by
\begin{align}
r_{s,1}^{DF}=&\frac{1}{2}\left[r^{DF}_1-\log\left(1+|h_e|^2\alpha\bar{P}\right)\right]^+, \label{eq_sec_rate_dec_1} \\
r_{s,2}^{DF}=&\frac{1}{2}\left[r^{DF}_2-\log\left(1+\frac{|h_e|^2\bar{\alpha}\bar{P}}{1+|h_e|^2\alpha \bar{P}}\right)\right]^+, \label{eq_sec_rate_dec_2}
\end{align}
where the extra multiplication by $\frac{1}{2}$ is due to transmission of the same message over two phases with equal durations.

Now that we settled the achievable secrecy rates, we turn to further optimizing the beamforming vector ${\bm d}_o$. Toward that end, we rewrite the constraint in (\ref{eq_dec_null}) slightly differently as follows:
\begin{align} \label{eq_dec_null_alt}
{\bm d}_o=\mathcal{P}^\perp\!\left({\bm g}_e\right){\bm u}_d,
\end{align}
where $\mathcal{P}^\perp\!\left(\cdot\right)$, as defined in (\ref{eq_p_perp_original}), now represents a projection matrix onto the orthogonal complement of vectors in $\mathbb{C}^K$, and ${\bm u}_d\in\mathbb{C}^K$ is some vector to be designed. Next, it is direct to see that $r^{DF}_1$ is non-decreasing in $\left|{\bm g}_1^\dagger{\bm d}_o\right|^2$ and that $r^{DF}_2$, after simple first derivative analysis, is also non-decreasing in $\left|{\bm g}_2^\dagger{\bm d}_o\right|^2$. Hence, one needs to choose ${\bm d}_o$ to maximize these terms. We propose maximizing their convex combination $\beta\left|{\bm g}_1^\dagger{\bm d}_o\right|^2+(1-\beta)\left|{\bm g}_2^\dagger{\bm d}_o\right|^2$, for some $0\leq\beta\leq1$ of choice. Using (\ref{eq_dec_null_alt}), and after simple manipulations, the optimal $\hat{{\bm u}}_d$ that maximizes such convex combination is given by the solution of the following problem:
\begin{align} \label{opt_dec_beta}
\max_{{\bm u}_d}\quad&{\bm u}_d^\dagger\mathcal{P}^\perp\!\left({\bm g}_e\right)\left(\beta{\bm g}_1{\bm g}_1^\dagger+(1-\beta){\bm g}_2{\bm g}_2^\dagger\right)\mathcal{P}^\perp\!\left({\bm g}_e\right){\bm u}_d \nonumber \\
\mbox{s.t.}\quad&{\bm u}_d^\dagger\mathcal{P}^\perp\!\left({\bm g}_e\right){\bm u}_d=1,
\end{align}
and therefore $\hat{{\bm u}}_d$ is given by the leading eigenvector of the (Hermitian) matrix:
\begin{align}
\mathcal{P}^\perp\!\left({\bm g}_e\right)\left(\beta{\bm g}_1{\bm g}_1^\dagger+(1-\beta){\bm g}_2{\bm g}_2^\dagger\right)\mathcal{P}^\perp\!\left({\bm g}_e\right),
\end{align}
i.e., the eigenvector corresponding to its largest eigenvalue. Finally, the optimal $\hat{{\bm d}}_o$ that solves problem (\ref{opt_dec_beta}) is given by
\begin{align}
\hat{{\bm d}}_o=\frac{\mathcal{P}^\perp\!\left({\bm g}_e\right)\hat{{\bm u}}_d}{\|\mathcal{P}^\perp\!\left({\bm g}_e\right)\hat{{\bm u}}_d\|}.
\end{align}
Given $\hat{{\bm d}}_o$, we substitute in (\ref{eq_sec_rate_dec_1}) and (\ref{eq_sec_rate_dec_2}) to get the achievable secrecy rates.

\subsection{Amplify-and-Forward} \label{sec_trst_amp}

In this subsection, we discuss the amplify-and-forward scheme. As in the decode-and-forward scheme, communication takes place over two phases and the eavesdropper overhears the communication during the two phases.

In the first phase, the received signals at the legitimate users, eavesdropper, and relays are given by (\ref{eq_rec_sig_1})--(\ref{eq_rec_sig_2}), (\ref{eq_rec_sig_e}), and (\ref{eq_rec_sig_r}), respectively, with a total transmit power $\bar{P}\leq P$. In the second phase, the $k$th relay amplifies its received signal by multiplying it by a constant $a_k$ and sends it through the channel. Effectively, this can be written as the multiplication: $\texttt{diag}\left({\bm a}\right){\bm y}_r$, where ${\bm a}\in\mathbb{C}^K$ is a beamforming vector to be designed, and $\texttt{diag}\left({\bm a}\right)$ is a diagonalization of the vector ${\bm a}$. The received signals at the legitimate users and the eavesdropper in the second phase are given by
\begin{align}
y^r_1&={\bm g}_1^\dagger\texttt{diag}\left({\bm a}\right){\bm y}_r+n^r_1, \\
y^r_2&={\bm g}_2^\dagger\texttt{diag}\left({\bm a}\right){\bm y}_r+n^r_2, \\
y^r_e&={\bm g}_e^\dagger\texttt{diag}\left({\bm a}\right){\bm y}_r+n^r_e.
\end{align}
Now observe that from, e.g., the strong user's perspective, this amplify-and-forward scheme can be viewed, using (\ref{eq_rec_sig_r}), as the following SIMO system:
\begin{align}
\begin{bmatrix}y_1\\y^r_1\end{bmatrix}=\begin{bmatrix}h_1\\{\bm g}_1^\dagger\texttt{diag}\left({\bm a}\right){\bm h}_r\end{bmatrix}x+\begin{bmatrix}n_1\\\tilde{n}^r_1\end{bmatrix},
\end{align}
where the noise term $\tilde{n}^r_1\triangleq{\bm g}_1^\dagger\texttt{diag}\left({\bm a}\right){\bm n}_r+n^r_1$ is complex-Gaussian with zero mean and variance $\mathbb{E}\left[|\tilde{n}^r_1|^2\right]={\bm g}_1^\dagger\texttt{diag}\left({\bm a}^*\right)\texttt{diag}\left({\bm a}\right){\bm g}_1+1$, with the superscript $*$ denoting the conjugate operation. One can write similar equations for the weak user as well. Hence, the achievable rates at the legitimate users of this SIMO system after the second phase are given by \cite[Section 5.3.1]{tse-wireless}
\begin{align}
r^{AF}_1&\!=\!\log\!\left(\!1\!+|h_1|^2\alpha\bar{P}+\frac{{\bm a}^\dagger{\bm G}_{1,r}{\bm a}}{1+{\bm a}^\dagger{\bm G}_1{\bm a}}\alpha\bar{P}\!\right), \label{eq_rate_amp_1} \\
r^{AF}_2&\!=\!\log\!\left(\!1\!+\frac{|h_2|^2\bar{\alpha}\bar{P}}{1+|h_2|^2\alpha\bar{P}}+\frac{{\bm a}^\dagger{\bm G}_{2,r}{\bm a}\bar{\alpha}\bar{P}}{1+{\bm a}^\dagger{\bm G}_2{\bm a}+{\bm a}^\dagger{\bm G}_{2,r}{\bm a}\alpha\bar{P}}\!\right), \label{eq_rate_amp_2}
\end{align}
where the superscript $AF$ is to denote amplify-and-forward achievable rates, and
\begin{align}
{\bm G}_{j,r}&\triangleq\texttt{diag}\left({\bm h}_r^*\right){\bm g}_j{\bm g}_j^\dagger\texttt{diag}\left({\bm h}_r\right),\quad j=1,2, \\
{\bm G}_j&\triangleq\texttt{diag}\left({\bm g}_j^*\right)\texttt{diag}\left({\bm g}_j\right),\quad j=1,2.
\end{align}
Since $|g_{1,k}|^2\geq|g_{2,k}|^2$, $1\leq k\leq K$, it can be shown that $\frac{{\bm a}^\dagger{\bm G}_{1,r}{\bm a}\bar{\alpha}\bar{P}}{1+{\bm a}^\dagger{\bm G}_1{\bm a}+{\bm a}^\dagger{\bm G}_{1,r}{\bm a}\alpha\bar{P}}\geq\frac{{\bm a}^\dagger{\bm G}_{2,r}{\bm a}\bar{\alpha}\bar{P}}{1+{\bm a}^\dagger{\bm G}_2{\bm a}+{\bm a}^\dagger{\bm G}_{2,r}{\bm a}\alpha\bar{P}}$, $\forall {\bm a}\in\mathbb{C}^K$, and hence, successive interference cancellation would be successfully employed at the strong user.

As for the eavesdropper, observe that by (\ref{eq_rec_sig_r}) we have
\begin{align}
{\bm g}_e^\dagger\texttt{diag}\left({\bm a}\right){\bm y}_r={\bm g}_e^\dagger\texttt{diag}\left({\bm a}\right){\bm h}_rx+{\bm g}_e^\dagger\texttt{diag}\left({\bm a}\right){\bm n}_r.
\end{align}
Upon noting that ${\bm g}_e^\dagger\texttt{diag}\left({\bm a}\right){\bm h}_r={\bm g}_e^\dagger\texttt{diag}\left({\bm h}_r\right){\bm a}$, we propose, for $K\geq2$, designing the beamforming vector ${\bm a}$ to be orthogonal to the vector $\texttt{diag}\left({\bm h}_r^*\right){\bm g}_e$ and denote it by ${\bm a}_o$. This way, the eavesdropper does not gain any useful information during the second phase. Thus, we have
\begin{align} \label{eq_amp_null}
{\bm g}_e^\dagger\texttt{diag}\left({\bm h}_r\right){\bm a}_o=0.
\end{align}
As in the decode-and-forward scheme, we further have for $K\geq3$ that $\{{\bm g}_1,~{\bm g}_2,~{\bm g}_e\}$ are linearly independent w.p. $1$, and therefore
\begin{align}
|{\bm g}_1^\dagger\texttt{diag}\left({\bm h}_r\right){\bm a}_o|>0,~|{\bm g}_2^\dagger\texttt{diag}\left({\bm h}_r\right){\bm a}_o|>0,\quad\text{w.p. $1$}.
\end{align}
Thus, the achievable secrecy rates in this case are given by
\begin{align}
r_{s,1}^{AF}=&\frac{1}{2}\left[r^{AF}_1-\log\left(1+|h_e|^2\alpha\bar{P}\right)\right]^+, \label{eq_sec_rate_amp_1} \\
r_{s,2}^{AF}=&\frac{1}{2}\left[r^{AF}_2-\log\left(1+\frac{|h_e|^2\left(1-\alpha\right)\bar{P}}{1+|h_e|^2\alpha \bar{P}}\right)\right]^+, \label{eq_sec_rate_amp_2}
\end{align}
where the extra multiplication by $\frac{1}{2}$ is due to transmission of the same message over two phases of equal durations, as in the decode-and-forward scheme.

We now focus on further optimizing the beamforming vector ${\bm a}_o$. Toward that end, we first note that the power transmitted in the second phase by the relays is given by
\begin{align}
&\mathbb{E}\left[{\bm a}_o^\dagger\texttt{diag}\left({\bm y}_r^*\right)\texttt{diag}\left({\bm y}_r\right){\bm a}_o\right] \nonumber \\
&\hspace{.05in}={\bm a}_o^\dagger\left(\texttt{diag}\left({\bm h}_r^*\right)\texttt{diag}\left({\bm h}_r\right)\bar{P}+{\bm I}_K\right){\bm a}_o\triangleq{\bm a}_o^\dagger{\bm A}{\bm a}_o. \label{eq_amp_pwr}
\end{align}
Next, we rewrite the constraint (\ref{eq_amp_null}) slightly differently as
\begin{align} \label{eq_amp_null_alt}
{\bm a}_o=\mathcal{P}^\perp\!\left(\texttt{diag}\left({\bm h}_r\right){\bm g}_e\right){\bm u}_a\triangleq{\bm F}{\bm u}_a
\end{align}
for some vector ${\bm u}_a\in\mathbb{C}^K$ to be designed. Next, we note that for the strong user, using (\ref{eq_amp_pwr}) and (\ref{eq_amp_null_alt}), finding the optimal ${\bm u}_a$ is tantamount to solving the following problem (note that ${\bm F}$ is a Hermitian matrix):
\begin{align}
\max_{{\bm u}_a}\quad&\frac{{\bm u}_a^\dagger{\bm F}{\bm G}_{1,r}{\bm F}{\bm u}_a}{1+{\bm u}_a^\dagger{\bm F}{\bm G}_1{\bm F}{\bm u}_a} \nonumber \\
\mbox{s.t.}\quad&{\bm u}_a^\dagger{\bm F}{\bm A}{\bm F}{\bm u}_a\!=\!P-\bar{P},
\end{align}
which can be equivalently rewritten as the following problem:
\begin{align}
\max_{{\bm u}_a}\quad&\frac{{\bm u}_a^\dagger{\bm F}{\bm G}_{1,r}{\bm F}{\bm u}_a}{{\bm u}_a^\dagger{\bm F}\left(\frac{1}{P-\bar{P}}{\bm A}+{\bm G}_1\right){\bm F}{\bm u}_a},
\end{align}
whose solution is given by the leading {\it generalized} eigenvector \cite{mtrx-comp} of the following matrix pencil:
\begin{align}
\left({\bm F}{\bm G}_{1,r}{\bm F}~,~{\bm F}\left(\frac{1}{P-\bar{P}}{\bm A}+{\bm G}_1\right){\bm F}\right),
\end{align}
i.e., the generalized eigenvector corresponding to the largest generalized eigenvalue of the pencil. Let us denote such vector by ${\bm u}_a^{(1)}$. Similarly, one can show that the optimal ${\bm u}_a$ for the weak user is given by the leading generalized eigenvector of the following matrix pencil:
\begin{align}
\hspace{-.1in}\left(\!{\bm F}{\bm G}_{2,r}{\bm F}~,~{\bm F}\left(\frac{1}{P-\bar{P}}{\bm A}+{\bm G}_2+{\bm G}_{2,r}\alpha\bar{P}\right){\bm F}\!\right),
\end{align}
which we denote by ${\bm u}_a^{(2)}$. To satisfy the power constraint (\ref{eq_amp_pwr}) and the orthogonality constraint (\ref{eq_amp_null_alt}), the corresponding ${\bm a}_o^{(j)}$, $j=1,2$, is given by
\begin{align}
{\bm a}_o^{(j)}=\sqrt{\frac{P-\bar{P}}{{\bm u}_a^{(j)T}{\bm F}{\bm A}{\bm F}{\bm u}_a^{(j)}}}{\bm F}{\bm u}_a^{(j)},\quad j=1,2.
\end{align}
Then, as in the decode-and-forward scheme, we propose choosing the optimal $\hat{{\bm a}}_o$ by the following convex combination:
\begin{align} \label{opt_amp_beta}
\hat{{\bm a}}_o=\beta{\bm a}_o^{(1)}+(1-\beta){\bm a}_o^{(2)}
\end{align}
for some $0\leq\beta\leq1$ of choice. Given $\hat{{\bm a}}_o$, we substitute in (\ref{eq_sec_rate_amp_1}) and (\ref{eq_sec_rate_amp_2}) to get the achievable secrecy rates.

%================================
\section{Communicating with an Untrusted Relay} \label{sec_untrstd}

In this section, we consider the situation in which an {\it untrusted} half-duplex relay node is available to assist with the BS's transmission.\footnote{We work with only one untrusted relay in this section, as opposed to the previous one, so as to better illustrate the main ideas behind the relaying schemes. The schemes, however, can be readily extended to the case of multiple untrusted relays by following, e.g., an opportunistic/worst case approach in which the design takes into consideration only the most effective/powerful relay node.} The relay is untrusted in the sense that it should be kept ignorant of the messages sent towards the users. However, it is assumed that the relay is {\it unmalicious} in the sense that it would not deviate from its transmission scheme, or attempt to hurt the users; it can only be curious enough to attempt to decode the users' messages. Such communication scenario has practical applications. For instance, users of a data providing network may have access to different data contents based on their subscription plans, or have hierarchal security clearances for different types of data. Since users are valid members of the same network, they have the incentive (or are required) to help each other and abide by the network protocols. The untrusted user, which is the relay in our case, however, can be curious enough to decode the contents of its received signals before forwarding them to the users. Such untrusted relay node is often called honest-but-curious in the literature, and has been adopted as the main model of study in, e.g., \cite{he-yener-two-hop-untrusted, zewail-untrusted-2hop-multi-receiver, he-yener-untrusted-relay}.

Let the received signal by the relay from the BS be given by
\begin{align}
y_r=h_rx+n_r,
\end{align}
where $h_r$ denotes the channel between the BS and the relay, and the noise term $n_r\sim\mathcal{C}\mathcal{N}(0,1)$. One direct approach to deal with this untrusted relay situation is to simply treat it as an external eavesdropper. This way, for a given $0\leq\alpha\leq1$, the following secrecy rates are achievable for this multi-receiver wiretap channel \cite[Theorem 5]{ersen-mr-wt} (these are the same as in (\ref{eq_sec_rate_1}) and (\ref{eq_sec_rate_2}) after replacing $h_e$ with $h_r$):
\begin{align}
r_{s,1}&=\left[\log\left(1+|h_1|^2\alpha P\right)-\log\left(1+|h_r|^2\alpha P\right)\right]^+, \\
r_{s,2}&=\left[\log\left(\!1+\frac{|h_2|^2\bar{\alpha}P}{1+|h_2|^2\alpha P}\!\right)\!-\!\log\left(\!1+\frac{|h_r|^2\bar{\alpha}P}{1+|h_r|^2\alpha P}\!\right)\right]^+\!\!\!.\end{align}
Clearly, this leads to zero secrecy rates if the relay is closer to the BS than the users and has a relatively better channel, i.e., if $|h_r|^2>|h_1|^2$ (and hence $|h_r|^2>|h_2|^2$ by assumption). However, it has been shown in \cite{he-yener-untrusted-relay} that positive secrecy rates can be achieved via {\it compress-and-forward} and {\it amplify-and-forward} relaying schemes, as opposed to treating the relay as an external eavesdropper, when the relay-to-users channel is orthogonal to the BS-to-relay channel, which is the case for instance if the relay's operation is half-duplex as in this section.

In the sequel, we extend the ideas of \cite{he-yener-untrusted-relay} to work in the context of NOMA, i.e., with multiple receivers, under two different modes of operation, namely, the {\it passive user} mode and the {\it active user} mode, as discussed next.

\subsection{Passive User Mode}

\begin{figure}[t]
\center
\includegraphics[scale=.9]{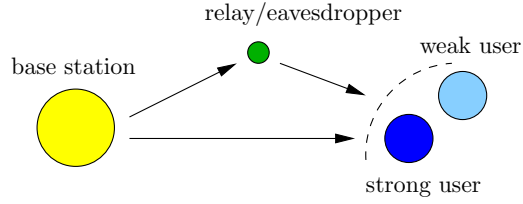}
\caption{Downlink NOMA system model with an untrusted relay node, and {\it passive} users.}
\label{fig_sys_model_passive}
\end{figure}

In the passive user mode, communication occurs over two phases. During the first phase, the BS broadcasts its messages to the users and to the relay. Then, the relay employs either a compress-and-forward or an amplify-and-forward scheme during the second phase to transmit its received message in the first phase towards the users, see Fig.~\ref{fig_sys_model_passive}. The received signals at the users during the second phase are given by
\begin{align}
y^r_1&=g_1x_r+n^r_1, \\
y^r_2&=g_2x_r+n^r_2,
\end{align}
where $x_r$ is the signal transmitted by the relay, $g_j$ is the channel between the relay and user $j$, and the noise terms $n_{j,r}$, $j=1,2$, are i.i.d. $\sim\mathcal{C}\mathcal{N}(0,1)$. The relay-to-users channel gains are such that $|g_1|^2\geq|g_2|^2$, i.e., the strong user with respect to the BS is also strong with respect to the untrusted relay, as assumed in the trusted relays scenario of Section~\ref{sec_trstd}. The system's total power budget $P$ is divided into $\bar{P}\leq P$ for the BS and $P-\bar{P}$ for the relay. We discuss the achievable secrecy rates under the two relaying schemes next.

\subsubsection{Compress-and-Forward} \label{sec-cf-passive}

Under the compress-and-forward scheme, the relay compresses its received signal $y_r$ into another signal $\hat{y}_r\triangleq y_r+n_Q$, where $n_Q$ is the quantization noise, and then encodes the quantized signal into its transmitted signal $x_r$. Following the results in \cite[Theorem 3]{he-yener-untrusted-relay}, setting $n_Q\sim\mathcal{C}\mathcal{N}\left(0,\sigma_Q^2\right)$ and $x_r\sim\mathcal{C}\mathcal{N}\left(0,P-\bar{P}\right)$, the achievable rates at the users under superposition coding are given by
\begin{align}
r_1^{CF,P}&=I\left(x;h_1x+n_1,h_rx+n_r+n_Q|s_2\right) \nonumber \\
&=\log\left(1+|h_1|^2\alpha \bar{P}+\frac{|h_r|^2\alpha \bar{P}}{1+\sigma_Q^2}\right), \label{eq_cf_1} \\
r_2^{CF,P}&=I\left(s_2;h_2x+n_2,h_rx+n_r+n_Q\right) \nonumber \\
&=\log\left(1+\frac{|h_2|^2\bar{\alpha}\bar{P}}{1+|h_2|^2\alpha \bar{P}}+\frac{|h_r|^2\bar{\alpha}\bar{P}}{1+|h_r|^2\alpha \bar{P}+\sigma_Q^2}\right), \label{eq_cf_2}
\end{align}
where the superscript $CF,P$ is to denote the compress-and-forward scheme under the passive user mode, and $I(\cdot;\cdot)$ denotes the mutual information measure \cite{cover}. The quantization noise power is designed to ensure decodability at both users by satisfying \cite[Theorem 3]{he-yener-untrusted-relay}
\begin{align}
I\left(x_r;g_1x_r+n^r_1\right)\!>\!I\left(h_rx+n_r+n_Q;h_rx+n_r|h_1x+n_1\right),
\end{align}
i.e.,
\begin{align} \label{eq_cf_var_q_1}
\log\left(1+|g_1|^2(P-\bar{P})\right)\!>\!\log\left(1+\frac{\left(|h_r|^2+|h_1|^2\right)\bar{P}+1}{\left(|h_1|^2\bar{P}+1\right)\sigma_Q^2}\right),
\end{align}
for the strong user, and
\begin{align}
I\left(x_r;g_2x_r+n^r_2\right)\!>\!I\left(h_rx+n_r+n_Q;h_rx+n_r|h_2x+n_2\right),
\end{align}
i.e.,
\begin{align} \label{eq_cf_var_q_2}
\log\left(1+|g_2|^2(P-\bar{P})\right)\!>\!\log\left(1+\frac{\left(|h_r|^2+|h_2|^2\right)\bar{P}+1}{\left(|h_2|^2\bar{P}+1\right)\sigma_Q^2}\right),
\end{align}
for the weak user. Upon observing that the achievable rates in (\ref{eq_cf_1}) and (\ref{eq_cf_2}) are both decreasing in $\sigma_Q^2$, we get from the above inequalities that one should set
\begin{align}
\sigma_Q^2&=\max\left\{\frac{\left(|h_r|^2+|h_1|^2\right)\bar{P}+1}{|g_1|^2(P-\bar{P})(|h_1|^2\bar{P}+1)},\frac{\left(|h_r|^2+|h_2|^2\right)\bar{P}+1}{|g_2|^2(P-\bar{P})(|h_2|^2\bar{P}+1)}\right\} \nonumber \\
&=\frac{\left(|h_r|^2+|h_2|^2\right)\bar{P}+1}{|g_2|^2(P-\bar{P})(|h_2|^2\bar{P}+1)},
\end{align}
where the second equality follows since $|h_1|^2\geq|h_2|^2$ and $|g_1|^2\geq|g_2|^2$. Therefore, the achievable secrecy rates are given by \cite[Theorem 3]{he-yener-untrusted-relay}
\begin{align}
r_{s,1}^{CF,P}&=\frac{1}{2}\left[r_1^{CF,P}-\log\left(1+|h_r|^2\alpha \bar{P}\right)\right]^+, \label{eq_psv_cf_1} \\
r_{s,2}^{CF,P}&=\frac{1}{2}\left[r_2^{CF,P}-\log\left(1+\frac{|h_r|^2\bar{\alpha}\bar{P}}{1+|h_r|^2\alpha \bar{P}}\right)\right]^+, \label{eq_psv_cf_2}
\end{align}
where the extra $1/2$ term is due to sending the same message over two phases of equal durations as in Sections~\ref{sec_dec_fwd} and \ref{sec_trst_amp}.

\subsubsection{Amplify-and-Forward} \label{sec-af-passive}

Under the amplify-and-forward scheme, the relay multiplies its received signal by a factor $\beta$ and forwards it to the users. Hence, one can treat the overall system as a SIMO system from the users' view point, as done in Section~\ref{sec_trst_amp}, and therefore the achievable rates at the users under superposition coding are given by
\begin{align}
r_1^{AF,P}&=\log\left(1+|h_1|^2\alpha \bar{P}+\frac{|g_1|^2\beta^2|h_r|^2\alpha \bar{P}}{1+|g_1|^2\beta^2}\right), \\
r_2^{AF,P}&\!=\!\log\!\left(\!1+\frac{|h_2|^2\bar{\alpha}\bar{P}}{1\!+\!|h_2|^2\alpha \bar{P}}+\frac{|g_2|^2\beta^2|h_r|^2\bar{\alpha}\bar{P}}{1\!+\!|g_2|^2\beta^2\left(1\!+\!|h_r|^2\alpha \bar{P}\right)}\!\right),
\end{align}
where the superscript $AF,P$ is to denote the amplify-and-forward scheme under the passive user mode, and the term $\beta$ satisfies the following power constraint:
\begin{align}
\beta^2=\frac{P-\bar{P}}{1+|h_r|^2\bar{P}}.
\end{align}
Note that since $|g_1|^2\geq|g_2|^2$, it holds that $\frac{|g_1|^2\beta^2|h_r|^2\bar{\alpha}\bar{P}}{1\!+\!|g_1|^2\beta^2\left(1\!+\!|h_r|^2\alpha \bar{P}\right)}\geq\frac{|g_2|^2\beta^2|h_r|^2\bar{\alpha}\bar{P}}{1\!+\!|g_2|^2\beta^2\left(1\!+\!|h_r|^2\alpha \bar{P}\right)}$, ensuring successful employment of successive interference cancellation at the strong user. Now the achievable secrecy rates are given by
\begin{align}
r_{s,1}^{AF,P}&=\frac{1}{2}\left[r_1^{AF,P}-\log\left(1+|h_r|^2\alpha \bar{P}\right)\right]^+, \label{eq_psv_af_1} \\
r_{s,2}^{AF,P}&=\frac{1}{2}\left[r_2^{AF,P}-\log\left(1+\frac{|h_r|^2\bar{\alpha}\bar{P}}{1+|h_r|^2\alpha \bar{P}}\right)\right]^+, \label{eq_psv_af_2}
\end{align}
where, again, the extra $1/2$ term is due to sending the same message over two equal phases.

\subsection{Active User Mode} \label{sec_active}

\begin{figure}[t]
\center
\includegraphics[scale=.9]{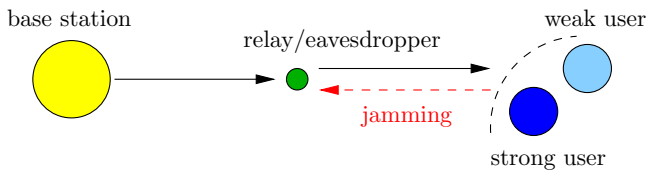}
\caption{Downlink NOMA system model with an untrusted relay node, and {\it active} users.}
\label{fig_sys_model_active}
\end{figure}

In the active user mode, communication also occurs over two phases as in the passive user mode except that the users send a {\it cooperative jamming} signal during the first phase to confuse the relay, and hence the notation {\it active user}. Our focus is on half-duplex nodes, and therefore we assume that the users cannot receive the BS's signal during the first phase while they are sending the jamming signal; instead, they only rely on the signal received from the relay during the second (forwarding) phase to decode their messages. Thus, in effect, there is no direct link between the BS and the users in the active user mode, and the model converts to a two-hop network, see Fig.~\ref{fig_sys_model_active}.

Let ${\bm J}z$ denote the jamming signal, with the beamfoming vector ${\bm J}\in\mathbb{C}^2$ and $z\sim\mathcal{CN}(0,1)$. Thus, the received signal at the relay during the first phase is now given by
\begin{align}
y_r=h_rx+{\bm g}^\dagger{\bm J}z+n_r,
\end{align}
where ${\bm g}\triangleq[g_1~g_2]$. The system's total power budget in this case is divided into $\bar{P}\leq P$ for the BS, $P-\bar{P}-\delta$ for the relay, and $\delta\leq P-\bar{P}$ for the users.

In order to maximally diminish the relay's decoding ability, the beamforming vector is chosen as
\begin{align}
{\bm J}=\frac{{\bm g}}{\|{\bm g}\|}\sqrt{\delta}.
\end{align}
The beamforming vector ${\bm J}$ is computed at the relay and then shared with the two users so that they compute their cooperative jamming signal.\footnote{Observe that the specific design of the beamforming vector in this case merely requires sharing the relay-strong user's channel gain with the weak user and vice versa.} We discuss the achievable secrecy rates under the two relaying schemes, compress-and-forward and amplify-and-forward, next.

\subsubsection{Compress-and-Forward}

The compress-and-forward scheme with active users is similar to the scheme presented in Section~\ref{sec-cf-passive} except that there is no direct link. In addition, the users subtract their jamming signal from their received signals from the relay before decoding. Hence, the achievable rates at the (active) users are given by
\begin{align}
r_1^{CF,A}&=\log\left(1+\frac{|h_r|^2\alpha \bar{P}}{1+\sigma_Q^2}\right), \label{eq_r1_cf_a} \\
r_2^{CF,A}&=\log\left(1+\frac{|h_r|^2\bar{\alpha}\bar{P}}{1+|h_r|^2\alpha \bar{P}+\sigma_Q^2}\right), \label{eq_r2_cf_a}
\end{align}
where the superscript $CF,A$ is to denote the compress-and-forward scheme under the active user mode, and the quantization noise power $\sigma_Q^2$ satisfies the same inequalities in (\ref{eq_cf_var_q_1}) and (\ref{eq_cf_var_q_2}) after setting the direct links' gains $h_1=h_2=0$ and replacing $P-\bar{P}$ by $P-\bar{P}-\delta$. Hence, upon recalling that $|g_1|^2\geq|g_2|^2$, $\sigma_Q^2$ is now given by
\begin{align} \label{eq_sigma_q_active}
&\sigma_Q^2=\frac{|h_r|^2\bar{P}+1}{|g_2|^2(P-\bar{P}-\delta)}.
%&{\color{blue}\sigma_Q^2}=\min\left\{\frac{\left(|h_r|^2+|h_1|^2\right)\bar{P}+1}{|g_1|^2(P-\bar{P}-\delta)(|h_1|^2\bar{P}+1)},\right. \nonumber \\
%&\left.\hspace{1in}\frac{\left(|h_r|^2+|h_2|^2\right)\bar{P}+1}{|g_2|^2(P-\bar{P}-\delta)(|h_2|^2\bar{P}+1)}\right\}.
\end{align}
Therefore, the achievable secrecy rates are given by
\begin{align}
r_{s,1}^{CF,A}&=\frac{1}{2}\left[r_1^{CF,A}-\log\left(1+\frac{|h_r|^2\alpha \bar{P}}{1+\|{\bm g}\|^2\delta}\right)\right]^+, \label{eq_actv_cf_1} \\
r_{s,2}^{CF,A}&=\frac{1}{2}\left[r_2^{CF,A}-\log\left(1+\frac{|h_r|^2\bar{\alpha}\bar{P}}{1+\|{\bm g}\|^2\delta+|h_r|^2\alpha \bar{P}}\right)\right]^+. \label{eq_actv_cf_2}
\end{align}

\subsubsection{Amplify-and-Forward}

Proceeding similarly as above, the users subtract their jamming signal from their received signals from the relay before decoding. This can be done if the term $\beta$ is known at the users, which is achieved by sharing the BS-to-relay channel gain $h_r$, along with the relay's transmit power, with them. Following the approach in Section~\ref{sec-af-passive}, the achievable rates at the (active) users under the amplify-and-forward scheme are given by
\begin{align}
r_1^{AF,A}&=\log\left(1+\frac{|g_1|^2\beta^2|h_r|^2\alpha \bar{P}}{1+|g_1|^2\beta^2}\right), \label{eq_r1_af_a} \\
r_2^{AF,A}&=\log\left(1+\frac{|g_2|^2\beta^2|h_r|^2\bar{\alpha}\bar{P}}{1+|g_2|^2\beta^2\left(1+|h_r|^2\alpha \bar{P}\right)}\right), \label{eq_r2_af_a}
\end{align}
where the term $\beta$ now satisfies the following power constraint:
\begin{align} \label{eq_beta_active}
\beta^2=\frac{P-\bar{P}-\delta}{1+|h_r|^2\bar{P}}.
\end{align}
Therefore, the achievable secrecy rates in this case are given by
\begin{align}
r_{s,1}^{AF,A}&=\frac{1}{2}\left[r_1^{AF,A}-\log\left(1+\frac{|h_r|^2\alpha \bar{P}}{1+\|{\bm g}\|^2\delta}\right)\right]^+, \label{eq_actv_af_1} \\
r_{s,2}^{AF,A}&=\frac{1}{2}\left[r_2^{AF,A}-\log\left(1+\frac{|h_r|^2\bar{\alpha}\bar{P}}{1+\|{\bm g}\|^2\delta+|h_r|^2\alpha \bar{P}}\right)\right]^+. \label{eq_actv_af_2}
\end{align}

\begin{remark}
We note that, operationally, the case in which the users do not transmit a cooperative jamming signal is equivalent to the passive user mode. Mathematically, however, setting the cooperative jamming power $\delta=0$ in (\ref{eq_actv_cf_1})-(\ref{eq_actv_cf_2}) and (\ref{eq_actv_af_1})-(\ref{eq_actv_af_2}) does \emph{not} yield back the secrecy rates achieved in the passive user mode given by (\ref{eq_psv_cf_1})-(\ref{eq_psv_cf_2}) and (\ref{eq_psv_af_1})-(\ref{eq_psv_af_2}), respectively. The reason behind this is that in the active user mode, the considered system is a two-hop network with no direct link, and setting $\delta=0$ does \emph{not} physically re-establish such direct link. The original passive user mode secrecy rates cannot be achieved by merely setting $\delta=0$ then, but by rather changing the mode of operation from the beginning. We also note that comparing the performances of passive and active user modes is not straightforward; we discuss this in more detail in Section~\ref{sec_num} below.
%Note that the case of $\delta=0$ is operationally equivalent to the users being passive, and hence the direct link from the BS to the users is reestablished. We therefore proceed with the assumption that $\delta>0$. Otherwise, the users would stay silent for half of the communication session unnecessarily. Comparing the performance of passive and active modes is not straightforward though. For instance, as we show below, setting $\delta=0$ in the rates achieved for active users, while mathematically approvable, does {\it not} give the same rates achieved by passive users. This is mainly because in the active user mode, the system becomes a two-hop network with no direct link from the BS to the users, unlike in the passive user mode. We discuss this in more detail and compare the performance of passive and active users in Section~\ref{sec_num}.
\end{remark}

\begin{remark} \label{rmrk_actv}
Upon substituting $\sigma_Q^2$ of (\ref{eq_sigma_q_active}) in (\ref{eq_r1_cf_a}) and (\ref{eq_r2_cf_a}), and substituting $\beta^2$ of (\ref{eq_beta_active}) in (\ref{eq_r1_af_a}) and (\ref{eq_r2_af_a}), we readily get that $r_1^{CF,A}<r_1^{AF,A}$ and $r_2^{CF,A}=r_2^{AF,A}$, making the amplify-and-forward scheme more useful in the active user mode (with respect to the strong user) than the compress-and-froward scheme. This is mainly attributed to the fact that in the second phase of compress-and-forward, the relay designs the quantization noise so that both users are able to decode, and hence the weak user's channel \emph{dominates}, unlike the amplify-and-forward scheme that allows each user to relatively make the best use of its channel. This comparison does not follow, however, in the passive user mode. In there, the presence of a direct channel from the BS to the legitimate users allows either scheme (compress-and-forward or amplify-and-forward) to potentially outperform the other, depending on the CSI and other system parameters, which we elaborate on in Section~\ref{sec_num}.
\end{remark}

%================================
\section{Numerical Evaluations} \label{sec_num}

In this section, we present some numerical examples to assess the performance of the proposed schemes in this paper. We first discuss how to evaluate the optimal power budget distribution among the communicating nodes, such that the achievable secrecy rate region is maximized. Specifically, we characterize the boundary of the achievable secrecy rate region by solving the following problem:
\begin{align} \label{opt_main}
\max_{\alpha,\bar{P}}\quad&\mu r_{s,1}^n+(1-\mu)r_{s,2}^n \nonumber \\
\mbox{s.t.}\quad&0\leq\bar{P}\leq P,\quad0\leq\alpha\leq1
\end{align}
for some $\mu\in[0,1]$, and the superscript $n$ differentiates between the proposed schemes in the paper, i.e., $n$ can be $J$, $DF$, or $AF$ for the trusted relays scenario of Section~\ref{sec_trstd}, or it can be $(CF,P)$, $(AF,P)$, $(CF,A)$, or $(AF,A)$ for the untrusted relay scenario of Section~\ref{sec_untrstd}. When considering the active user mode in the untrusted relay scenario discussed in Section~\ref{sec_active}, we also maximize over the jamming power $\delta$, under the constraint: $\delta\leq P-\bar{P}$. We use a line search algorithm to numerically solve the above problem. Note that the feasible set is bounded, which facilitates the convergence of the algorithm to an optimal point. We set $\beta=\mu$ in (\ref{opt_dec_beta}) and (\ref{opt_amp_beta}) so as to design the beamforming vector in proportion to the priority given to each user in the weighted sum rate maximization problem (\ref{opt_main}).

The physical layout that we consider is a simple one-dimensional system, where the strong user is located at $30$ meters away from the BS, the weak user at $40$ meters away, and the eavesdropper at $50$ meters away. For the trusted relays scenario, we have $K=5$ relays, and for simplicity we assume that they are all close enough to each other that they are approximately at the same distance of $15$ meters away from the BS. To emphasize the effect of distance on the channel gains, we use the following simplified channel model \cite{petropulu-coop-relay-security}: $h=\sqrt{1/l^\gamma}e^{j\theta}$, where $h$ is the channel gain between two nodes, $l$ is the distance between them, $\gamma=3.5$ is the path loss exponent, and $\theta$ is a uniform random variable in $[0,2\pi]$. We denote by $l_1$, $l_2$, $l_e$, and $l_r$ the distances from the BS to the strong user, weak user, eavesdropper, and relays, respectively. We set $P$ to $90$ dBm.\footnote{Note that the noise power is normalized in this paper, and hence $P$ also represents the SNR.} We run $1000$ iterations of the simulations and compute the average performance.

\begin{figure}[t]
\center
\includegraphics[scale=.45]{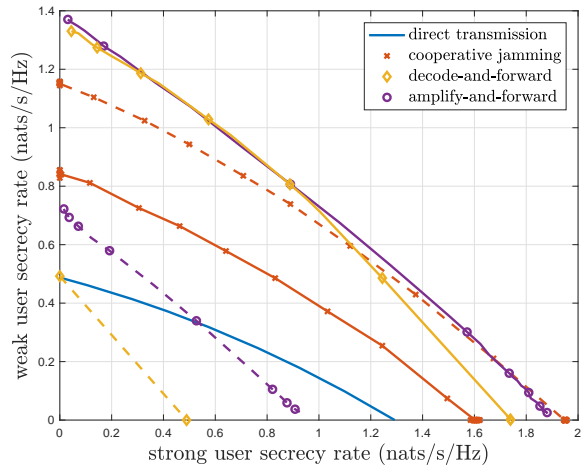}
\caption{Achievable secrecy rate regions of the proposed schemes for the trusted relays scenario. Solid lines are when $l_e=50$ meters, and dashed lines are when $l_e=20$ meters.}
\label{fig_sec_reg_all_schemes}
\end{figure}

We start by presenting some results for the trusted relays scenario of Section~\ref{sec_trstd}. In Fig.~\ref{fig_sec_reg_all_schemes}, we plot the achievable secrecy rate regions of the proposed schemes (cooperative jamming, decode-and-forward and amplify-and-forward) along with the direct transmission scheme (without the relays' help). Solid lines represent the system parameters stated above, and dashed lines represent the situation in which $l_e=20$ meters. We see that when $l_e=50$ meters, cooperative jamming outperforms direct transmission and is in turn outperformed by both decode-and-forward and amplify-and-forward which perform relatively close to each other. With $l_e=20$ meters, direct transmission achieves zero secrecy rates since the eavesdropper is closer to the BS than both users. However, strictly positive secrecy rates are achievable by all the relaying schemes. We also see that cooperative jamming performs best in this case, since the relays are close to the eavesdropper and hence their jamming effect is quite powerful. For parts of the region, it even performs very close to decode-and-forward and amplify-and-forward for the $l_e=50$ meters case.

\begin{figure}[t]
\center
\includegraphics[scale=.45]{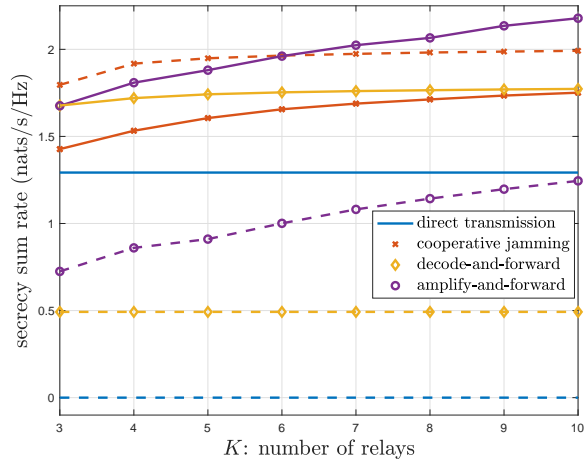}
\caption{Achievable secrecy sum rates of the proposed schemes for the trusted relays scenario vs. the number of relays. Solid lines are when $l_e=50$ meters, and dashed lines are when $l_e=20$ meters.}
\label{fig_sum_sec_rate_nmbr_relays}
\end{figure}

Next, we show the effect of the number of relays on the achievable {\it secrecy sum rates} of the proposed schemes in Fig.~\ref{fig_sum_sec_rate_nmbr_relays}. Again we observe that all relaying schemes achieve positive secrecy sum rate when $l_e=20$ meters and that cooperative jamming performs best in this case. We also observe that amplify-and-forward is more sensitive than the other schemes to the number of relays, and that the decode-and-forward scheme's performance does not change much with the number of relays, especially for the case when $l_e=20$ meters. This is primarily due to the fact that in (\ref{eq_dec_rate_1}) and (\ref{eq_dec_rate_2}), the larger the number of relays, the larger the first term in the minimum regarding the relays-to-users rate gets, yet the second term in the minimum regarding the BS-to-relays rate stays almost the same and represents the bottleneck to the system.

\begin{figure}[t]
\center
\includegraphics[scale=.45]{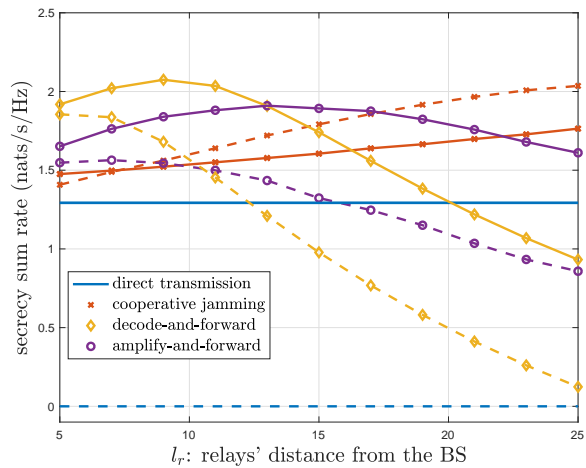}
\caption{Achievable secrecy sum rates for the trusted relays scenario vs. the relays' distance from the BS. Solid lines are when $l_e=50$ meters, and dashed lines are when $l_e=27$ meters.}
\label{fig_sum_sec_rate_relays_distance}
\end{figure}

Finally, we fix the users' distances and show the effect of the relays' distance from the BS (and hence the users) on the achievable secrecy sum rates in Fig.~\ref{fig_sum_sec_rate_relays_distance}. For this case, we vary the relays' distance but still keep them closer to the BS than the legitimate users and the eavesdropper. The dashed lines in this case are when $l_e=27$ meters. We see that only the cooperative jamming scheme's performance monotonically increases with the relays' distance from the BS, which is again attributed to the fact that the jamming effect is more powerful when the relays get closer to the eavesdropper. For both the decode-and-forward and amplify-and-forward schemes, their performance is best when the relays are midway between the BS and the users, with decode-and-forward performing better for smaller values of $l_r$ than amplify-and-forward, and vice versa.

\begin{figure}[t]
\center
\includegraphics[scale=.45]{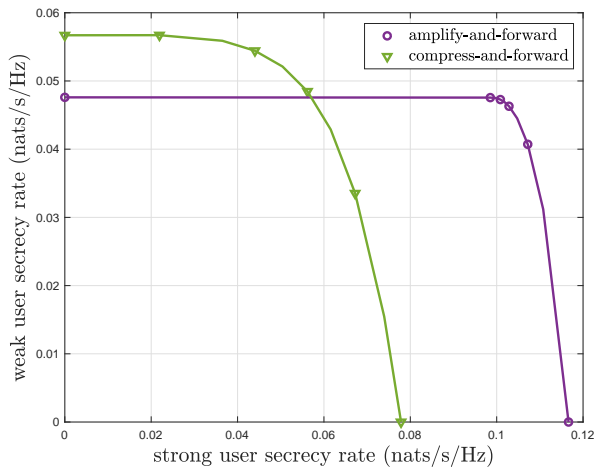}
\caption{Achievable secrecy rate region for the untrusted relay scenario with {\it passive} users.}
\label{fig_passive_reg}
\end{figure}

\begin{figure}[t]
\center
\includegraphics[scale=.45]{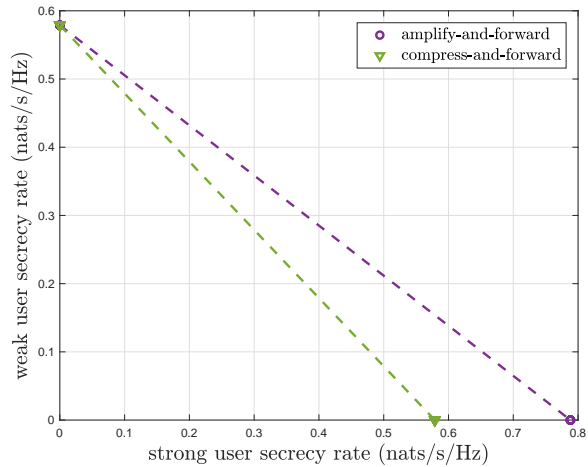}
\caption{Achievable secrecy rate region for the untrusted relay scenario with {\it active} users.}
\label{fig_active_reg}
\end{figure}

We now conclude this section by presenting some results for the untrusted relay scenario of Section~\ref{sec_untrstd}. For this case, we slightly change the distances to $l_1=40$ meters, $l_2=50$ meters, and $l_r=30$ meters. Note that treating the relay as an eavesdropper achieves zero secrecy rates in this case, since the relay is closer to the BS than both users. Thus, we only show the compress-and-forward and the amplify-and-forward achievable secrecy rates for this setting. In Fig.~\ref{fig_passive_reg}, we plot the achievable secrecy rate regions of the proposed schemes under the passive user mode. We see that amplify-and-forward performs better for the strong user, while compress-and-forward performs better for the weak user. Choosing the best relaying scheme therefore depends on which part of the region the system is to operate at. In Fig.~\ref{fig_active_reg}, we plot the achievable secrecy rate regions of the proposed schemes under the active user mode. As mentioned in Remark~\ref{rmrk_actv}, we see in this case that amplify-and-forward outperforms compress-and-forward.

%In Fig.~\ref{fig_passive_reg}, we plot the achievable secrecy rate regions of the proposed schemes under the passive user mode, along with the direct transmission scheme in which the relay is treated as an eavesdropper. We also plot the results for the case when $l_r=55$ meters. We see that when $l_r=35$ meters, direct transmission achieves zero secrecy rates since the relay is closer to the BS than the users, while the two proposed schemes achieve strictly positive secrecy rates. When $l_r=55$ meters, the proposed schemes still outperform direct transmission for some parts of the region, which shows the benefit of using the help of the untrusted relay even if it is relatively further away from the BS. In both cases, when $l_r=35$ meters and when $l_r=55$ meters, compress-and-forward completely outperforms amplify-and-forward.

\begin{figure}[t]
\center
\includegraphics[scale=.45]{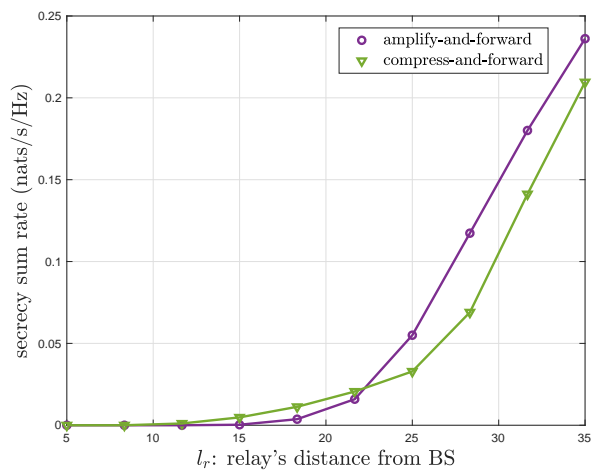}
\caption{Achievable secrecy sum rate for the untrusted relay scenario with {\it passive} users vs. the relay's distance from the BS.}
\label{fig_psv_dr}
\end{figure}

\begin{figure}[t]
\center
\includegraphics[scale=.45]{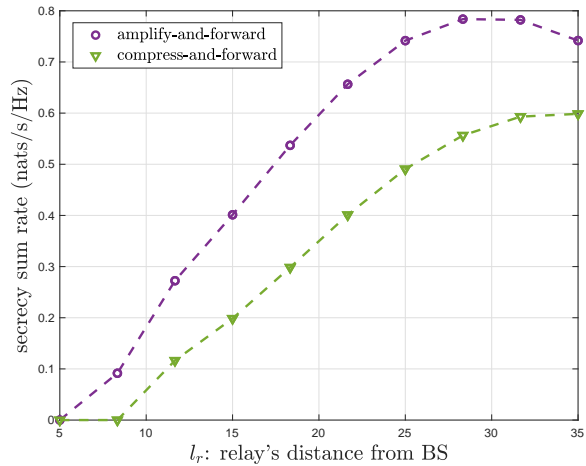}
\caption{Achievable secrecy sum rate for the untrusted relay scenario with {\it active} users vs. the relay's distance from the BS.}
\label{fig_actv_dr}
\end{figure}

Next, we show the effect of the relay's location, $l_r$, on the achievable secrecy sum rate, for the passive user mode in Fig.~\ref{fig_psv_dr}, and the active user mode in Fig.~\ref{fig_actv_dr}. From the figures, we see that under passive users, compress-and-forward performs better when the relay is relatively closer to the BS, and is outperformed by amplify-and-forward when the relays gets further away. Under active users, as we have seen in Fig.~\ref{fig_active_reg}, amplify-and-forward outperforms compress-and-forward for all values of $l_r$. We also see that the optimal value of $l_r$ that maximizes the secrecy sum rate achievable under compress-and-forward is {\it not} the furthest possible from the BS.

Finally, upon comparing Fig.~\ref{fig_passive_reg} to Fig.~\ref{fig_active_reg}, and Fig.~\ref{fig_psv_dr} to Fig.~\ref{fig_actv_dr}, we observe that the secrecy rates achieved under active users are generally better than those achieved under passive users.

In what follows, we present a summary of implications of the numerical results presented in this section in Figs.~\ref{fig_sec_reg_all_schemes} through~\ref{fig_actv_dr}, with the purpose of providing insights into secure system design of NOMA cooperative communications:
\begin{itemize}
\item When there is a close-by eavesdropper, and there exists a number of trusted relays, one should use cooperative jamming. On the other hand, when the eavesdropper is far away, one should focus on either decode-and-forward or amplify-and-forward.
\item The amplify-and-forward (resp. decode-and-forward) scheme is the most (resp. least) sensitive to the number of trusted relays in the network, when it comes to the rate of increase of the secrecy sum rate.
\item Whether one has a number of trusted relays, or needs to deal with an untrusted relay, the location of the relay(s) relative to the BS (and hence the users/eavesdropper) greatly affects the achievable secrecy rates, and needs to be carefully designed.
\item In the case of an untrusted relay, compress-and-forward can be more useful than amplify-and-forward only in the passive user mode. In the active user mode, amplify-and-forward performs better.
\item In general, active users achieve higher secrecy rates than passive users in the untrusted relay scenario.
\end{itemize}

%================================
\section{Conclusion and Future Directions}

Secure transmission schemes in a downlink two-user SISO NOMA system have been considered, under two main scenarios: existence of an external eavesdropper, and communicating through an untrusted half-duplex relay. For the first scenario, employment of trusted cooperative half-duplex relays has been proposed, and various relaying schemes have been studied, namely, cooperative jamming, decode-and-forward, and amplify-and-forward. For each scheme, secure beamforming signals have been derived at the relays to boost up the secrecy rates achieved at the users. For the second scenario, two modes of operation have been studied, namely, passive user mode and active user mode, under each of which two relaying schemes have been considered, namely, compress-and-forward and amplify-and-forward. For either of the two considered scenarios, the performance of each relaying scheme has been thoroughly analyzed and compared, under the same total system power budget. The results have shown that choosing the best relaying scheme and/or mode of operation is highly dependent on the physical layout of the network, especially the distances between the nodes. Extensive tradeoffs among the different schemes have also been discussed.

One direct step forward as a future direction for this line of research would be to extend the schemes developed in this paper to nodes with multiple antennas; the case with more than two legitimate users; and to the case without eavesdropper's CSI availability. Another direction would be to consider full-duplex relay nodes, in which case one would be able to combine, e.g., decode-and-forward simultaneously with cooperative jamming in the trusted cooperative relays scenario to further hurt the eavesdropper, similar to the approach followed in \cite{raef-df-cj}. It would also be of interest to extend the results of the untrusted relay scenario for multiple untrusted relays and/or add an external eavesdropper.


\end{document}